\documentclass[a4]{article}
\usepackage[dvips]{graphicx}

\begin{document}

\title{Towards a new technique of incoherent scatter signal processing.}

\author{O. I. Berngardt, B. G. Shpynev}

\date{}
\maketitle
{\par\centering{
Institute of Solar-Terrestrial Physics,

POBox 4026, Irkutsk, 664033, Russia

(berng@iszf.irk.ru)
}
\par}
\begin{abstract}
This paper offers a new technique of incoherent scattering signal processing.
The technique is based on the experimentally observed comb structure of the
spectral power of separate realizations. The technique implies determining the
positions and amplitudes of peaks in separate realizations, the formation -
on their basis - of the spectral power of an individual realization not distorted
by the smoothing function, and a subsequent summation of such spectra for the
realizations. The technique has been tested using data from the Irkutsk incoherent
scatter radar, both for the case of the incoherent scattering from thermal irregularities
of plasma and for the case of the aspect scattering from instabilities elongated
with the geomagnetic field.
\end{abstract}

%%\begin{article}

\section{Introduction.}

The incoherent scatter (IS) method is one of a number of ionospheric remote
sensing techniques. The method provides geophysical parameters of the ionosphere
over a wide height range (from 100 to 1000 km), with spatial and temporal resolution
determined by the form of the sounding pulse, and by the particular procedure
of processing the received signal [{\it Holt et al.}, 1992]. The accuracy of geophysical
parameters determination in this case is usually governed by many parameters:
ionospheric parameters, and parameters of the receiver (for instance, the background
noise temperature), the method of the received signal processing, the type of
sounding signal, the received signal averaging time, and by the spectral resolution
of the method (or by the delay resolution in the case of a correlational processing
of the signal). Furthermore, it is often necessary to improve the spectral resolution
without impairing the spatial resolution. A conventional approach in handling
this problem involves using special (`composite') signals, with a subsequent
special-purpose processing of the received signal 
[{\it Farley},1972; {\it Sulzer}, 1993; {\it Lehtinen},1986].
However, situations can at times arise where it is not appropriate to use composite
pulses (from energy considerations or auxillary conditions, for example), so
that it is necessary to have a technique which would work well with traditional
('simple') pulsed signals.

There are currently two main techniques for processing the received backscattered
signal: the correlational technique, and the spectral technique [{\it Evans}, 1963].
Since the correlation function (obtained by applying the former type of processing)
and the spectral power of the received signal (obtained by applying the latter
type of processing) are related by the Wiener-Khintchin theorem, the two types
of processing are equivalent in principle.

This paper is concerned with the method of improving the spectral resolution
of the incoherent scatter method, based on the properties of the received signal
according to the data from the Irkutsk incoherent scatter (IS) radar without
using composite sounding signals.

\section{The existing technique for processing the signal, and experimental data}

Let us consider the radar equation relating the spectral power of the scattered
signal to fluctuation parameters of dielectric permittivity. Within the approximation
of the single scattering, in the far zone of the antenna, the mean spectral
power of the received signal \( <|u(\omega )|^{2}> \) is defined by a statistical
radar equation [{\it Berngardt and Potekhin}, 2000] which - under the assumption of the spatial homogeneity
of the spectral density of the irregularities and its weak dependence on the
modulus of the wave vector \( \Phi (\nu ,\overrightarrow{r},-\overrightarrow{e_{r}}k)=\Phi (\nu ,\overrightarrow{e_{r}}r_{0}-\overrightarrow{e_{r}}2k_{0}) \)
reduces to:

\begin{equation}
\label{eq:1}
<|u(\omega )|^{2}>=\int S(\omega -\nu )<|u_{teor}(\nu )|^{2}>d\nu 
\end{equation}

\begin{equation}
\label{eq:2}
<|u_{teor}(\nu )|^{2}>=V_{0}\int \Phi (\nu ,\overrightarrow{e_{r}}r_{0},-\overrightarrow{e_{r}}2k_{0})|g(\overrightarrow{e_{r}})|^{2}d\overrightarrow{e_{r}}
\end{equation}

here \( \Phi (\nu ,\overrightarrow{r},\overrightarrow{k})=\int <\epsilon (t,\overrightarrow{r})\epsilon ^{*}(t+\Delta t,\overrightarrow{r}+\Delta \overrightarrow{r})>exp(-i\nu t+i\overrightarrow{k}\Delta \overrightarrow{r})\frac{d\nu d\overrightarrow{k}}{(2\pi )^{4}} \)
- is the steady-state spectral density of permittivity irregularities; \( g(\overrightarrow{e_{r}}) \)
is the beam factor determined by the product of the beams of the transmit and
receive antennas; \( \overrightarrow{e_{r}}=\overrightarrow{r}/r \) is a unit
vector in a given direction; \( V_{0} \) is the sounding volume; \( r_{0} \)
is the mean distance to it; \( k_{0} \) is the wave number of the sounding
wave; and \( <|u_{teor}(\omega )|^{2}> \) is the theoretical spectrum of backscattering
not distorted by the smearing function \( S(\omega ) \).

The problem of improving the spectral resolution in this case implies using
the deconvolution operation with the kernel \( S(\omega )\approx |\int a(\omega -\nu )o(\nu )d\nu |^{2} \)
determined by the form of the sounding signal and the receiving window and usually
having the property:

\[
\frac{dS(\omega )}{d\omega }=0;\, when\, \omega =0;\, \frac{d^{2}S(\omega =0)}{d\omega ^{2}}<0,\]

that is, having only one maximum at the zero frequency.

Thus, formally, we need to define the \( DECONV \) - deconvolution operation:

\begin{equation}
\label{eq:3}
<|u_{teor}(\nu )|^{2}>=<|u(\omega )|^{2}>\, DECONV\, S(\omega -\nu )
\end{equation}

To carry out such an operation we make use of the linearity of the averaging
operation:

\begin{equation}
\label{eq:4}
<|u_{teor}(\nu )|^{2}>=<|u(\omega )|^{2}\, DECONV\, S(\omega -\nu )>.
\end{equation}

Hence, to improve the spectral resolution, it is necessary to apply the deconvolution
operation to each realization of spectral power and accumulate result spectrums
over the realizations. Generally the problem (\ref{eq:4}) is not simpler compared
with the initial one (\ref{eq:3}); however, in the case of its simplified solution,
one may take advantage of the following experimental evidence of the structure
of spectra of separate realizations.

\begin{figure*}
\vskip1in 
\resizebox*{0.9\textwidth}{!}{
\includegraphics{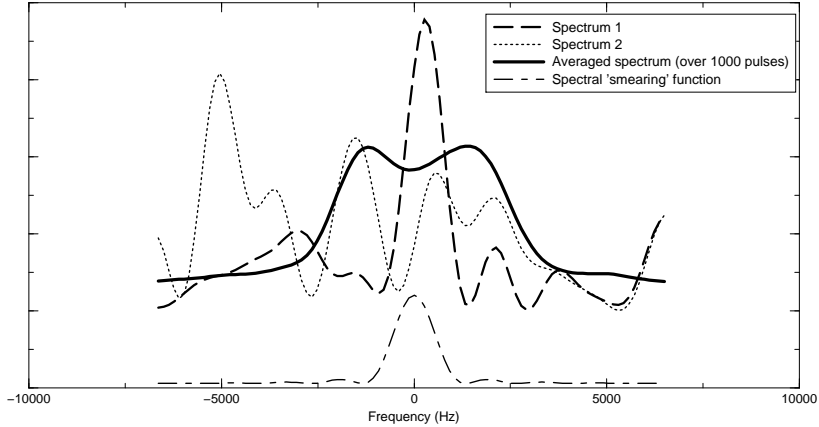}}
\caption{Form of the spectral power of two successive realizations of the incoherently
scattered signal (dotted line, and dashed line), mean spectral power (averaged
over 1000 realizations, thick solid line), spectral density of the 'smearing'
function (dash-dotted line).} 
\label{fig1}
\end{figure*}

Figure~\ref{fig1} presents the structure of spectral power of two successive realizations
of the scattered signal in sounding with the pulse of a duration of 750 ms and
the length of the receiving window of 750 ms. For comparison, the figure also
shows the form of the accumulated (averaged) spectral power, and the model form
of the 'smearing' function. Theoretically, the form of the smearing function
depends on a large number of ionospheric parameters (on the electron density
profile, on the experimental geometry with respect to the geomagnetic field),
and can differ rather strongly from the model form [{\it Shpynev}, 2000]; 
however,
the model form is applicable in the case of qualitative estimations.

It is apparent from the figure that the spectra of individual realizations differ
rather strongly from one another, which suggests that the processes are occurring
at a high rate when compared with the repetition frequency of sounding pulses,
and is in agreement with existing data (the lifetime of thermal irregularities
is on the order of 200 mks, which is significantly less than the interval between
separate sounding runs - at the Irkutsk IS radar it is about 40 ms).

\begin{figure*}
\vskip1in 
\resizebox*{0.9\textwidth}{!}{
\includegraphics{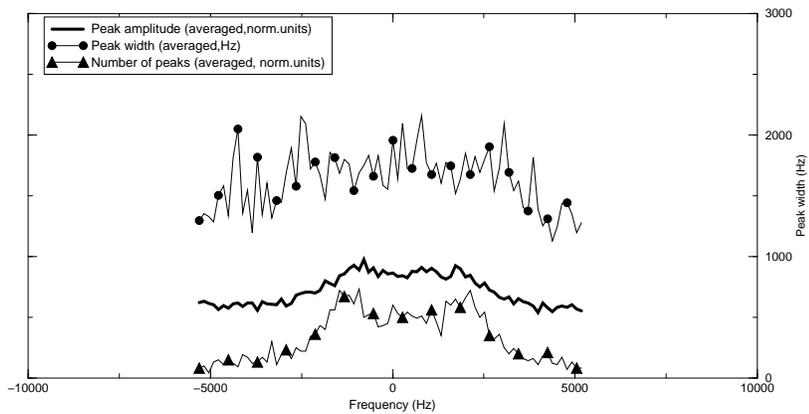}
}
\caption{Frequency dependence of the mean parameters of the spectral power comb
structure for single realization (averaging over 1000 realizations): mean amplitude
of the peaks (solid line, arbitrary units), mean width of the peaks (line with circles,
1000 Hz correspond to the spectral width of the sounding signal), and
the number of realizations having a peak at a given frequency (line with triangles,
arbitrary units).
} 
\label{fig2}
\end{figure*}

In spite of a relatively strong variability from realization to realization,
the fine 'comb' structure of the spectra is conserved. Consider the characteristics
of such a comb structure. Figure~\ref{fig2} presents the frequency dependence of the amplitude
of the peaks (solid line), the frequency dependence of the width of the peaks
(line with circles), and a total number of peaks at a given frequency for the
entire set of the realizations used in the analysis (line with triangles). It is evident
from Figure~\ref{fig2} that the width of the peaks varies within 1-2 spectral widths of
the sounding signal. The peaks are concentrated mainly in the band of the mean
spectral power of the received signal, and the amplitude and occurrence frequency
drop when the frequency of the peak is shifted with respect to the zero frequency.

\section{Technique for solving the problem - deconvolution before an averaging.}

The above characteristics of spectra of separate realizations of the received
signal suggest that the initial (not convoluted with the smearing function)
spectral power of the received signal has a comb structure with delta-shaped
combs. Furthermore, the width of the peaks in the experimentally measured spectrum
is determined solely by the properties of the smearing function. This permits
us to relatively easily perform a deconvolution in (\ref{eq:4}). Indeed, within
the framework of this assumption, a 'nonsmoothed' spectral power of a separate
realization has the form:\begin{equation}
\label{eq:5}
|u_{teor}(\omega )|^{2}=\begin{array}{c}
N\\
\sum \\
i=1
\end{array}A_{i}\delta (\omega -\omega _{i})
\end{equation}
Then, within a constant factor, we have\begin{equation}
\label{eq:6bis}
|u(\omega _{j})|^{2}=\begin{array}{c}
N\\
\sum \\
i=1
\end{array}A_{i}S(\omega _{j}-\omega _{i})
\end{equation}
We take into consideration that the function \( S(\omega ) \) is unknown but
it is sufficiently narrowbanded when compared with a total spectral width \( u(\omega ) \),
and the peaks in the spectral power of a separate realization are sufficiently
isolated from each other, which permits us to neglect the influence of one peak
on another. Then the amplitude of the observed peaks in \( |u(\omega _{j})|^{2} \)
will be proportional to the amplitude of the peaks in the 'nonsmeared' spectrum
\( |u_{teor}(\omega )|^{2} \), and theirs location are the same:\begin{equation}
\label{eq:6}
|u(\omega _{j})|^{2}=S(0)A_{j}+o(\omega _{j})
\end{equation}
where \( o(\omega _{j})=\begin{array}{c}
N\\
\sum \\
i=1
\end{array}A_{i}S(\omega _{j}-\omega _{i})-S(0)A_{j} \) is a small addition which - in the case of a sufficient separation of the peaks
(larger than the width of the smearing function \( S(\omega ) \))) becomes
zero. In accordance with the expression (\ref{eq:6}), we determine the amplitudes
\( A_{i} \) and the frequencies \( \omega _{i} \) from experimentally measured
spectra \( |u(\omega )|^{2} \), which corresponds to the solution of the system:

\begin{equation}
\label{eq:7}
\left\{ \begin{array}{c}
\left. \frac{d}{d\omega }|u(\omega )|^{2}\right| _{\omega =\omega _{i}}=0\\
\left. \frac{d^{2}}{d\omega ^{2}}|u(\omega )|^{2}\right| _{\omega =\omega _{i}}<0\\
A_{i}=|u(\omega _{i})|^{2}
\end{array}\right. 
\end{equation}

Upon determining in this way the set of pairs of the parameters \( A_{i} \)
and \( \omega _{i} \), it is also possible to obtain the 'nonsmeared' spectrum
of a single realization \( |u_{teor}(\omega )|^{2} \) by applying a deconvolution
in (\ref{eq:4}). If we exactly know the form of the smearing function \( S(\omega ) \),
the amplitudes \( A_{i} \) can be determined not in accordance with the last
equation of the system (\ref{eq:7}) but by solving a system of linear equations
for amplitudes with due regard for the form of the smearing function (\ref{eq:6bis}).

\section{Discussion of results}

The technique suggested here was used in processing the data on incoherent scattering
from thermal irregularities of ionospheric plasma using measurements from the
Irkutsk incoherent scatter radar. Because of the suggested technique (\ref{eq:5},\ref{eq:7})
was obtained for noice absence, the experimental data for testing was used with
high signal-to-noice ratio.

\begin{figure*}
\vskip1in 
\resizebox*{0.9\textwidth}{!}{
\includegraphics{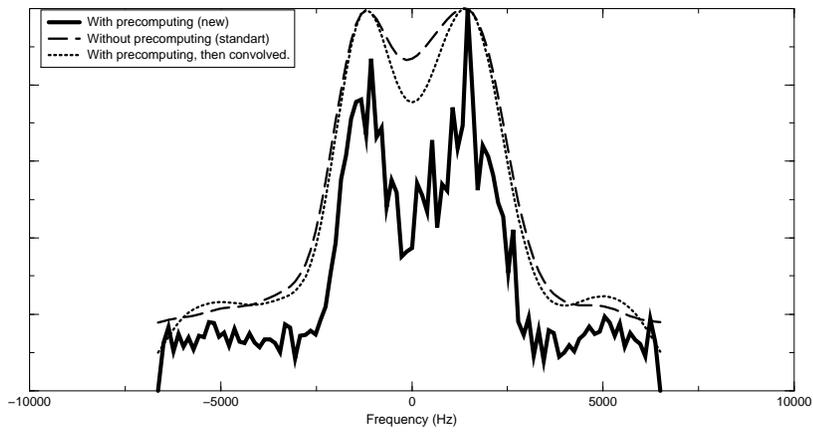}
}
\caption{Mean spectral power of the incoherently scattered signal (dashed line),
mean spectral power of the signal calculated by the proposed technique (thick
solid line) and calculated by the proposed technique, followed by a convolution
with the spectral smearing function (dotted line). Accumulation - 1000 realizations.
}
\label{fig3}
\end{figure*}

Figure~\ref{fig3} gives an example of a processing of incoherent scatter data in terms of
the model of (\ref{eq:5},\ref{eq:7}). The thick solid line corresponds to
a theoretical spectrum \( <|u_{teor}(\omega )|^{2}> \) calculated in terms
of this model; the dotted line shows the spectrum \( <|u(\omega )|^{2}> \)
measured by a traditional technique, and the thin line plots the function \( \int <|u_{teor}(\nu )|^{2}>S(\omega -\nu )d\nu  \)
which - in an ideal variant of the known \( S(\omega ) \) - must coincide with
\( <|u(\omega )|^{2}> \). Figure~\ref{fig3} clearly shows a good agreement between the
mean spectral power obtained as a result of a standard processing (dotted line)
and the spectrum with deconvolution which is convoluted with the theoretical
smearing function (thin solid line), which suggests a sufficiently good deconvolution
operation when calculating \( <|u_{teor}(\omega )|^{2}> \) by the algorithm
of (\ref{eq:5},\ref{eq:7}).

The relatively high dispersion level of the resulting spectrum \( <|u_{teor}(\omega )|^{2}> \)
(spectrum with high spectral resolution) is associated with the comb structure
of some of the realizations. Indeed, an actual averaging at a given frequency
occurs only over realizations involving a peak at a given frequency. Thus, to
an averaging over 1000 realizations there corresponds an averaging over about
70-100 real realizations involving a peak at a given frequency. It is evident
from Figure~\ref{fig3} that in the convolution with the smearing function this dispersion
decreases to a level similar to the weak dispersion level of a standard spectrum.
Since the inverse problem of obtaining physical parameters of the ionosphere
from the mean spectral power of the received signal is usually solved by 
fitting
the spectral form using the method of least squares [{\it Holt et al.}, 1992], such a dispersion should
not increase substantially the error of determining the parameters when compared
with standard processing techniques in the case of averaging over an equal number
of realizations.

\begin{figure*}
\vskip1in 
\resizebox*{0.9\textwidth}{!}{
\includegraphics{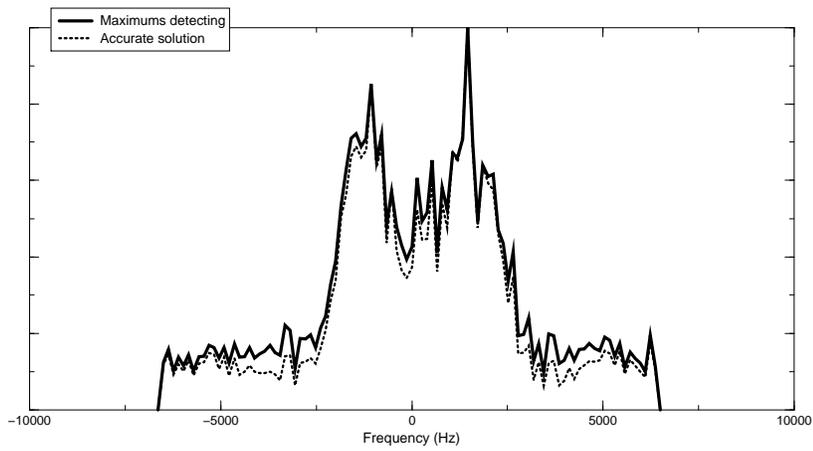}
}
\caption{
Form of the spectrum with the removal of the effect of the smearing
function by a direct solution of the system (\ref{eq:6bis}), with due regard
for the form of the smearing function (dotted line), and by a simple detection
of maxima (\ref{eq:7}) (thick line).
}
\label{fig4}
\end{figure*}

Taking into account the real 'smearing' function when calculating the amplitudes
\( A_{i} \) (which implies solving a system of linear equations for the amplitudes
\( A_{i} \) (\ref{eq:6bis})) does not give any perceptible decrease in the
dispersion of the spectrum, which suggests that the dispersion of spectra is
associated with inadequate accumulation. An example of a processing of spectra
following the proposed technique, which implies solving (\ref{eq:7}) and a
direct solution of the system (\ref{eq:6bis}), is given in 
Figure~\ref{fig4}.

\section{Using the technique in the real incoherent scattering experiment.}
The technique suggested has been used for incoherent scattering signal 
processing during November 15, 2001 experiment. All the data have obtained 
at Irlutsk Incoherent Scattering Radar(53N,103E). The radar operates in standard
regime with the following parameters:
Sounding frequency 158 MHz,
Pulse power 2.6MWt
Pulse duration 750mks
Pulse repeating frequency 24.4Hz
Antenna pattern main lobe is elongated with Earth magnetic field and 
nearly vertical.
The November 15, 2001 experiment is characterized by high electron density
in the main ionospheric maximum ( \( > 10^{6} cm^{-3} \) ). In the daytime the signal-to-noice 
relation exceeds 20, and the conditions are ideal for
analysis of the incoherent scattering spectra structural pecularities. In parallel 
with traditional processing of the IS signals, the row samples (realizations) have been 
recordered(approximately 1.4 GByte, for the 8:00LT-23:00LT period). These data
become a basis for the experimental comparison of the new technique with standard one.
The comaprison has been done by the following technique. For given altitude range
the incoherent scattering signal realizations (the pair of its quadrature components)
has been cutted and processed by the two different techniques: the standard one and 
the new one.

The first (standard technique) uses fast Fourier transform, accumulation of the 
obtained spectrums and using this averaged power spectrum as a source for the 
\( T_{e} \) and \( T_{i} \) calculation by the standard temperature calculation 
technique  [{\it Shpynev}, 2000]. The averaged power 
spectrum has been compared with model ones convolved with the smoothing function,
which depends on pulse duration and electron density.

The second (new technique) uses fast Fourier transform and deconvolving of the 
unaveraged spectrum with the smoothing function followed by the accumalation of 
the result. The result ('unconvolved' power spectrum) have been used as a source
for the \( T_{e} \) and \( T_{i} \) calculation by the standard temperature 
calucaltion technique [{\it Shpynev}, 2000], but it have 
been comparised with madel spectrum, without its convolving with smoothing function.

All the parameters (averaging time - 6 min) has been the same.

\begin{figure*}
\vskip0.5in 
\resizebox*{0.9\textwidth}{!}{
\includegraphics{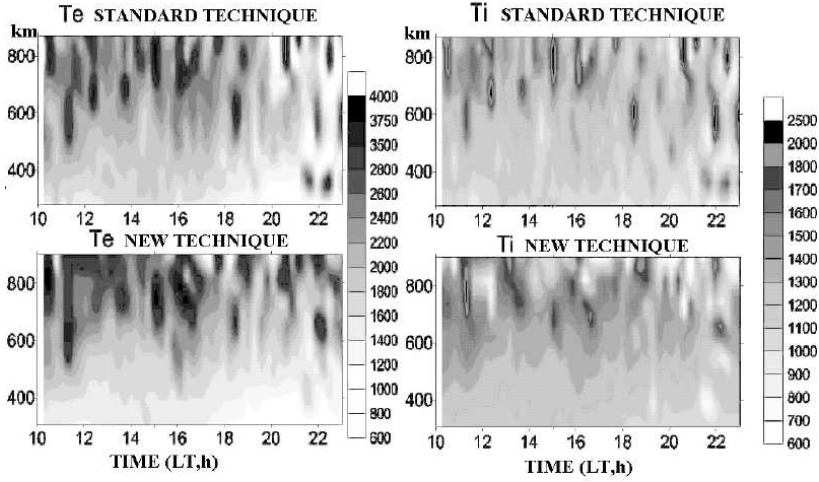}
}
\caption{
Electron and Ion temperatures during the November 15, 2001 experiment obtained 
with standard technique (top) and new technique (bottom).
}
\label{fig6}
\end{figure*}

\begin{figure*}
\vskip0.5in 
\resizebox*{0.9\textwidth}{!}{
\includegraphics{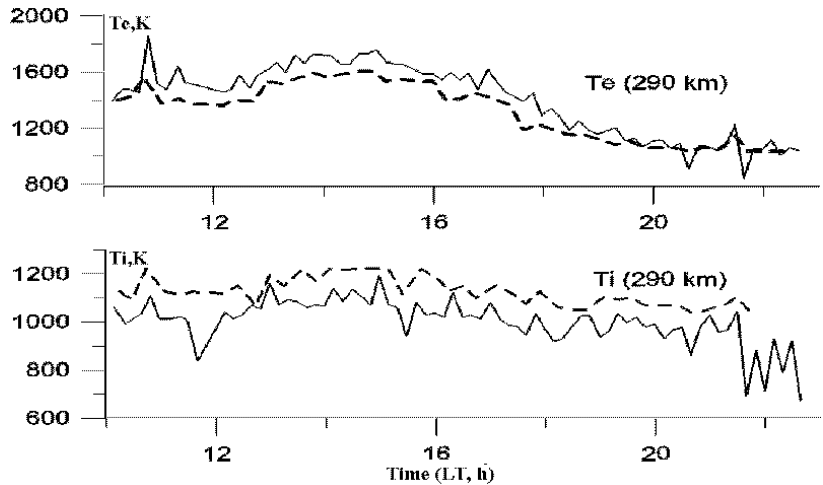}
}
\caption{
Electron and Ion temperatures during the November 15, 2001 experiment obtained 
with standard technique (solid line) and new technique (dashed line) at 290km.
}
\label{fig7}
\end{figure*}

The results are shown at Figure~\ref{fig6}. As one can see from comparison,
the difference between temperatures obtained differs slightly. For the new
technique the systematical decreasing of the \( T_{e} \) and
systematical increasing of the \( T_{i} \) are observed. You can see this clearly
from the Figure~\ref{fig7} (Temperatures for the 290km height). This error is 
within experiment error and comparison results could be stated as well. An 
another aspect of the obtained results is theirs smoother dependence on time - 
their time variations is smaller (as one can see from Figure~\ref{fig7} too).

\section{Conclusion}

A technique has been suggested for improving the spectral resolution in the
case of the incoherent scattering when sounding by simple (squared) pulses.
The method is based on an experimental model of single spectra of the scattered
signal realizations. Furthermore, the model of a single realization spectrum
represents a comb structure, the sum of delta-functions of a different amplitude
at different frequencies convoluted with a certain function that is determined
by the form of the sounding signal and of the receiving window. In terms of
such an empirical model, it is possible to perform a deconvolution of the mean
spectral power of the received signal with the smearing function in order to
obtain the mean spectral power, at the stage of analysis of separate realizations,
which is equivalent to an improvement of the spectral resolution of the method.
Furthermore, the model parameters are derived by a simple determination of the
positions and amplitudes of local maxima of the individual realizations spectra,
which does not require much computer time and can be implemented in the real-time
mode. Also, the technique does not require knowing the form of the 'smearing'
function and could therefore be used to improve the spectral resolution over
a wide range of cases.

\begin{figure*}
\vskip0.5in 
\resizebox*{0.9\textwidth}{!}{
\includegraphics{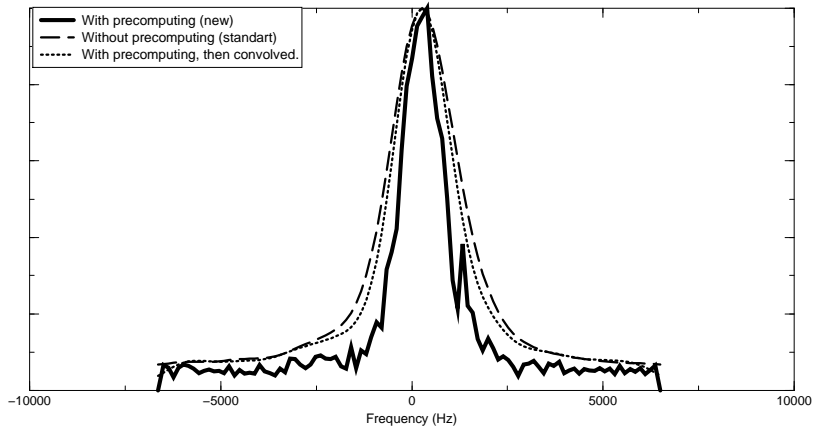}
}
\caption{
Mean spectral power of the backscattered signal from the E-layer irregularities
elongated with the geomagnetic field (dashed line), mean deconvolved spectral
power of the signal calculated by the proposed technique (thick solid line),
and calculated by the proposed technique with a subsequent convolution with
the spectral smearing function (dotted line). Accumulation - 1000 realizations.
}
\label{fig5}
\end{figure*}

The proposed technique appears to be able to be extended also to other types
of single backscattering from the ionosphere. As an example, Figure~\ref{fig5}
 illustrates
the processing of experimental data on another type of singly backscattered
signal from the ionosphere - backscattering from the E-layer irregularities
elongated with the geomagnetic field. All curves in this case are similar to
the curves in Figure~\ref{fig3} and, as in the case of 
Figure~\ref{fig3}, there is a good agreement
of \( <|u_{teor}(\omega )|^{2}> \) to the spectrum \( <|u(\omega )|^{2}> \)
with the removal of the convolution with the smearing function \( S(\omega ) \).

So, the proposed technique of incoherent scattered signal processing (\ref{eq:5},\ref{eq:7})
could be used in some cases for increasing the spectral resolution of the incoherent
scattering method. But, to use this technique one needs good signal to noice
ratio, and narrow enought spectral smoothing function to one could use approximation
(\ref{eq:7}) which does not depend on spectral smoothing function \( S(\omega ) \),
instead of accurate solution of (\ref{eq:6bis}).

\section*{Acknowledgments}
Authors are grateful to A.V.Medvedev for fruitful descussions. Work has been 
carried out under partial support of RFBR grants \#00-15-98509 and \#00-05-72026 .

%%\end{article}


\begin{thebibliography}{}
%%\bibitem[{\it Kofman} (1997)]{Kofman}
%%Kofman W., Plasma instabilities and their observations with the incoherent scatter
%%technique// Incoherent scatter: theory, practice and science, technical report
%%97/53 -EISCAT scientific association - 1997, p33-65.

\bibitem[{\it Berngardt and Potekhin} (2000)]{RLU}
Berngardt O.I. and Potekhin A.P., Radar equations in the radio wave backscattering
problem, {\it Radiophysics and Quantum electronics}, 43(6), 484--492, 2000

\bibitem[{\it Evans} (1963)]{Evans}
Evans J.V., Theory and practice of ionosphere study by Thomson scatter radar,
\it{Proc.IEEE}, 57, 496--530,1963

\bibitem[{\it Farley} (1972)]{Farley}
Farley D.T., Multiple-pulse incoherent scatter correlation function measurements,
{\it Radio Science}, 7(6), 661--666, 1972

\bibitem[{\it Holt et al.} (1992)]{Holt}
Holt J.M., Rhoda D.A., Tetenbaum D., van Eyken A.P.
Optimal analysis of incoherent scatter radar data, {\it Radio Science}, 27(3), 435--447, 1992

\bibitem[{\it Lentinen} (1986)]{Lehtinen}
Lehtinen M.S., Statistical theory of incoherent scatter radar measurements,
Ph.D.Thesis.- Univ. of Helsenki, -1986 -97p.

\bibitem[{\it Shpynev} (2000)]{Shpynev}
Shpynev B.G., Methods of Processing Incoherently Scattered Signals with the 
Fadaray Effect Taken Into Account., Ph.D.Thesis, Irkutsk, 2000, 142 p.(in Russian)

\bibitem[{\it Sulzer} (1993)]{Sulzer}
Sulzer M.P., A new type of alternating code for incoherent scatter measurements,
{\it Radio Science}, 28, 995--1001, 1993



\end{thebibliography}
\end{document}